\documentclass[aps,twocolumn,showpacs,superscriptaddress]{revtex4}
\usepackage{amsmath,amssymb}
\usepackage{graphics,graphicx}
\usepackage{dcolumn,bm}
\usepackage{psfrag}
\usepackage{color}
\newcommand{\cu}
{\affiliation{Department of Physics, University of Calcutta,
92 Acharya Prafulla Chandra Road, Kolkata 700009, India.}}

\begin{document}

\title{Virtual walks in spin space: a study in a family of two-parameter models}

\author{Pratik Mullick}%
\cu
\author{Parongama Sen}%
\cu

\normalsize  

\begin{abstract}

We investigate the dynamics of classical spins mapped as walkers in a virtual `spin' space using a generalised two-parameter family of spin models characterized by parameters $y$ and $z$ [M. J. de Oliveira, J. F. F. Mendes and M. A. Santos, J. Phys. A Math. Gen. \textbf{26}, 2317 (1993)]. The behavior of $S(x,t)$, the probability that the walker is at
position $x$ at time $t$ is studied in detail.  
In general  $S(x,t) \sim t^{-\alpha}f(x/t^{\alpha})$
with $\alpha \simeq 1$ or $0.5$ at large times depending on the parameters. 
In particular,  $S(x,t)$ for the point $y=1, z=0.5$ corresponding to the voter model shows a
crossover  in time;  associated with this crossover, two timescales can be
defined which vary with the system size $L$ as $L^2\log L$. 
We also show that as the voter model point is approached from  the disordered
regions along 
different directions, the width of  the Gaussian distribution $S(x,t)$ diverges in a power law manner with  different  exponents. 
For the majority voter case, the results indicate that the   
the virtual walk can detect the   phase  transition
 perhaps more efficiently compared to other non-equilibrium methods. 


\end{abstract}


\maketitle

\section{Introduction}

Non-equilibrium behavior associated with domain coarsening  phenomena in classical spin models below the critical temperature are usually
understood by studying the dynamics of  
domain growth, correlation functions, persistence probability etc. \cite{derrida1,derrida2,Bray1,Bray2}.  
For a long time it was believed that a single length scale (or time scale) governed the 
dynamics of the system: the domains  grow as $t^{1/z}$ and the same exponent $z$ 
dictates the dynamics of other relevant quantities like correlation function, number of spin flips, magnetization etc. 
Much later, the persistence probability $P(t)$, defined as the probability that a spin does not flip till 
time $t$ was shown to behave also as a power law. However, the associated exponent was found to be unrelated to other known exponents, either static or dynamic.
More recently, in the generalised voter model in two dimensions, the presence of two time scales
was found numerically \cite{parna} for a certain parameter range.
It is therefore interesting to extend the  studies in dynamical phenomena to see whether other time scales and
independent exponents exist. 

In certain cases it is possible to map the dynamics of coarsening to a different system, e.g.,  
in one dimension, the dynamics of the Ising model is equivalent to that of  interacting (annihilating) diffusive random walkers.
For the Voter model in any dimension, mapping to coalescing random walks is possible \cite{ligget}. 
In this paper we use the concept 
of a virtual walker which is related  to the dynamics of the spins    and call it a walk in the 
 the spin space, when the spin system undergoes a domain coarsening. Some aspects of this     walk has been studied 
in different systems earlier \cite{dornic_g98,drouffe98,newman,balda,luck}  and we explore its role   further  using a two parameter 
family of classical spin models in two dimensions. 
Virtual walks  have been considered  
in systems other than spin models also, e.g.    to study the stochastic properties, nucleotide sequences in a DNA
was mapped onto a walk \cite{peng}. For financial data, a random
walk picture was introduced as early as in 1900 by Louis Bachelier \cite{bachelier}. Later, similar walks
were studied for models of wealth exchange \cite{econo1,econo2}.

\begin{figure}
\includegraphics[width=6cm]{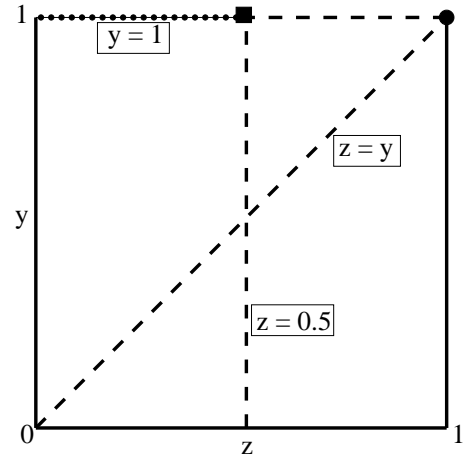}
 \caption{The $z-y$ parameter space showing the three lines for which the virtual walks were studied.
The dotted horizontal line indicates the region $y=1,0\leq z\leq 0.5$ and the dashed horizontal line indicates $y=1,0.5\leq z\leq 1$.
The dashed vertical line represents the region $z=0.5$, while the dashed inclined line represents the region $z=y$. The black square representing the point $z=0.5$, $y=1$ corresponds to the Voter model and the black circle representing
the point $z=1$, $y=1$ corresponds to the Ising model.}
\label{phase}
\end{figure}

In section II, we have described the models and the related walk in detail with a brief review of earlier works. The results obtained are
presented in section III and the analysis of the results are made in the following section.
In the last section, we present a summary of the work with some conclusive statements.

\section{The walk and the models}

We numerically simulate classical spin models on $L\times L$ square lattices and start with a completely random
 configuration. One Monte Carlo Step (MCS) comprises $L^2$ updates
and random asynchronous updating rule was used.
In each of these updates, a spin is chosen randomly and updated according to a certain rule;
the spin will either remain in the original state
or flip. 
To each of the spins in the lattice space we associate a walker in a so called virtual spin space (one dimensional),
which is allowed to move one step towards the `right' or `left' according to the following convention: 
the walker moves to the right (left) if the corresponding spin is in up (down) state after
the completion of one MCS. The displacement $x$ of the walker is in the virtual space; the initial
position is $x(t=0)=0$ by definition. It is easy to identify $x$ with the sum $\sum \sigma (t)$ where $\sigma = \pm 1$
are the two states of the spin variables.  
It is to be noted here that the real space position coordinates
of the spins in the lattice and $x$, the displacement of the walkers in the virtual space are uncorrelated.

The probability distribution $S(x,t)$ for the position
$x$ of the walkers at time $t$ is estimated. 
The range of $x$ is $-t \leq x \leq t$; the extreme values $x=\pm t$ correspond
to spins which have never flipped from the initial up/down state.
In fact $P(t)=S(-t,t)+S(t,t)$ and
because of up/down symmetry of the system, one can write
$S(\pm t,t)=\frac{1}{2}P(t)$.

We  briefly discuss  here the  earlier works in which the distribution $S(x,t)$ has been studied. In \cite{dornic_g98},
the distribution for Brownian motion, Ising model in one dimension and Voter model in two dimensions 
had been considered. As the distribution is related to the persistence property, several studies have been aimed at 
 obtaining  the  persistence exponent and related dynamical  quantities \cite{newman,balda,luck}. 
The nature of the distribution at different temperatures for the two dimensional Ising models has been studied to show that it changes as one crosses the  phase transition 
point \cite {drouffe2001}.   
In the present  work, our aim has been to extract as much information of the systems as possible from the 
nature of $S(x,t)$;  several  observations  which have not been  encountered earlier like crossover behaviour in time and related timescales, 
divergences related to the width of the distribution etc. are reported.   
The efficiency of detecting  a  phase transition using $S(x,t)$ compared to other methods has been also discussed.

The studies are restricted to the
non-equilibrium regime; this is because if the system reaches a steady state (either equilibrium or non-equilibrium), the walker will simply continue in the same direction forever.

We have considered a generalised two-parameter family of spin models with up-down symmetry on a
square lattice \cite{oliveira} to study the walk dynamics. The system  undergoes  single-spin flip stochastic dynamics.
The spin flip probability is given by
\begin{equation}
\omega_i(\sigma) = \frac{1}{2}[1 - \sigma_iF_i(\sigma)].
\end{equation}
Here, $\sigma_i=\pm1$ is the spin variable
at the $i$-th lattice site and $\sigma=\sum_{\delta}\sigma_{i+\delta}$ is the sum of the four neighbouring spins of $\sigma_i$.
The function $F_i(\sigma)$ is defined as $F(0) = 0$, $F(2) = -F(-2) = z$ and $F(4) = -F(-4) = y$, and the parameters $z$ and $y$ are restricted to
$z \leq 1$ and $y \leq 1$. 
We focus our studies on three different lines in the $z$-$y$ parameter
space (Fig. \ref{phase}): (i) $y = 1$, (ii) $z = y$ and (iii) $z = 0.5$. The points $z = 1$, $y = 1$ and $z=0.5$, $y=1$ correspond 
to the Ising model at zero temperature and Voter model respectively.\\

In general, we have attempted to fit $S(x,t)$ to a familiar scaling form:
\begin{equation} 
S(x,t) \sim t^{-\alpha}f(x/t^{\alpha}).
\label{scaling}
\end{equation}
Here $\alpha$ indicates the exponent relevant for the walk, e.g. for a diffusive walk, $\alpha=0.5$, while
for a ballistic walk, $\alpha=1$.

\begin{figure}
\includegraphics[width=8cm]{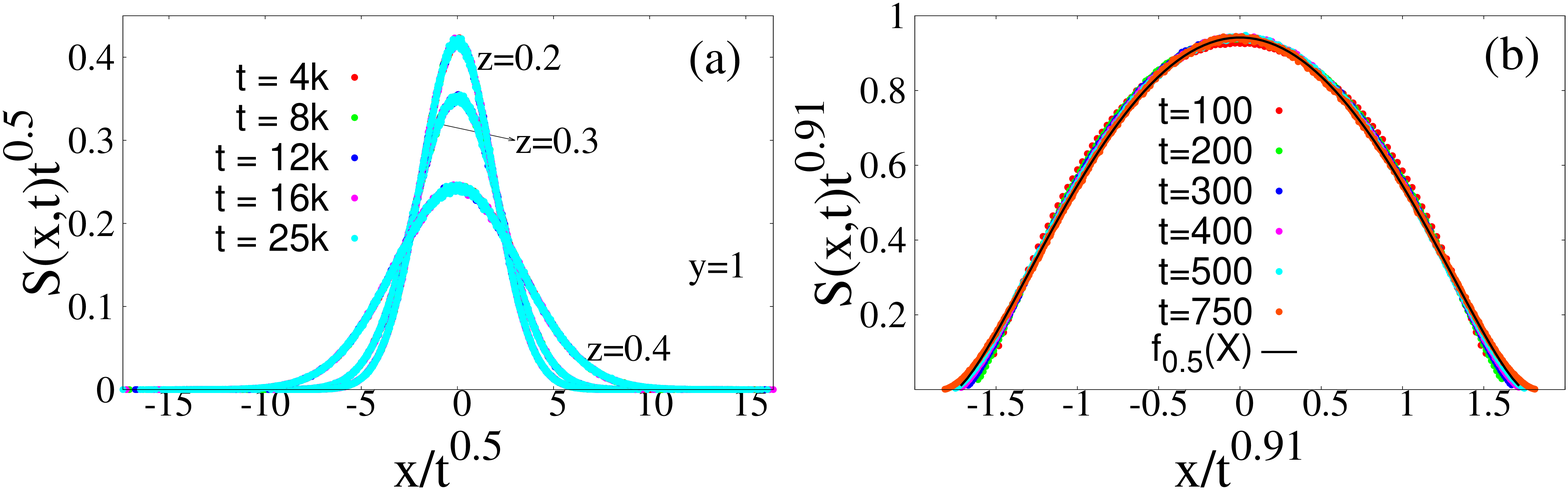}
 \caption{(a) Data collapse of $S(x,t)$ for $z=0.2,0.3,0.4$ and $y=1$ for a system size $L = 128$. For each value of $z$, the collapse was done for five different times. (b) Data collapse of $S(x,t)$ for $z=0.5,y=1$ for a system size $L=128$ for $t\leq 750$. The collapse was done for six different times. The collapsed data were fitted using the function $f_{0.5}(X)=0.201(1.75^2-X^2)^{1.38}$ (Eq. (\ref{f0.5})).}
\label{z0.5}
\end{figure}

\begin{figure}
\includegraphics[width=4cm]{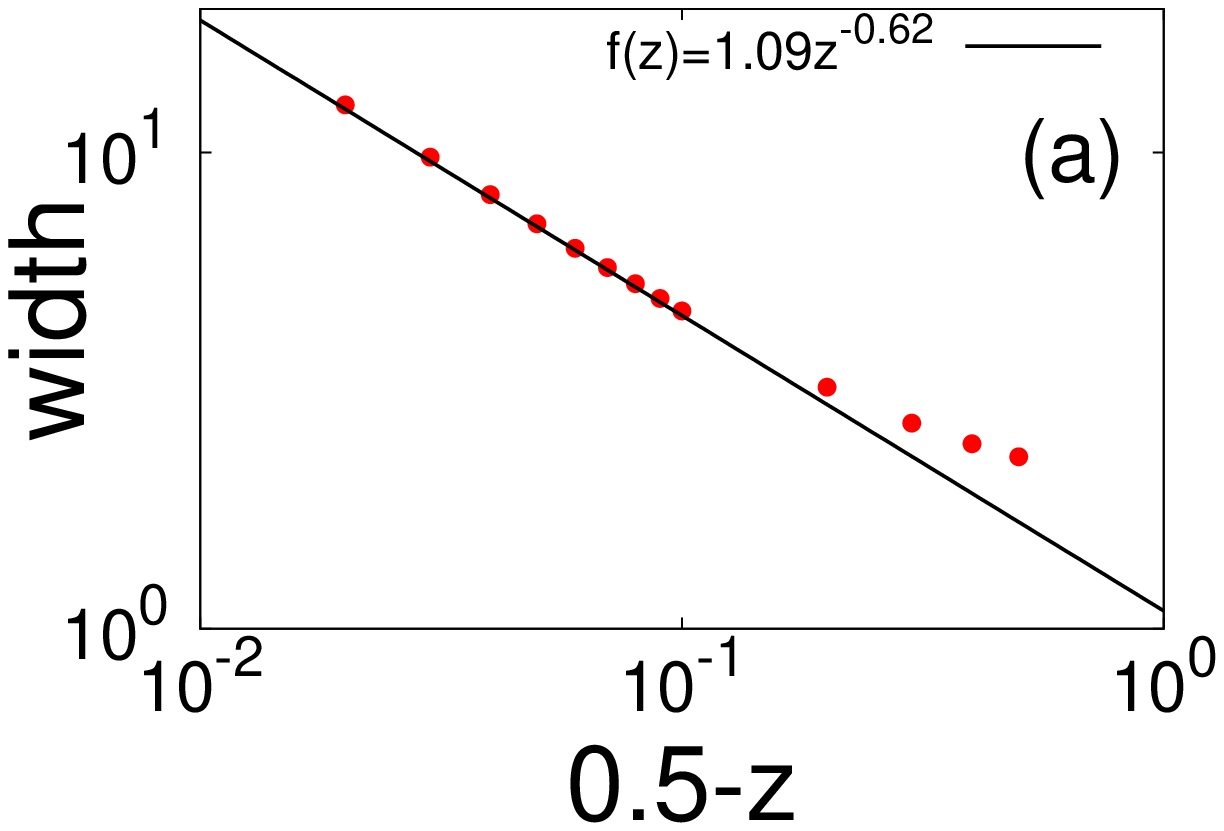}
\includegraphics[width=4cm]{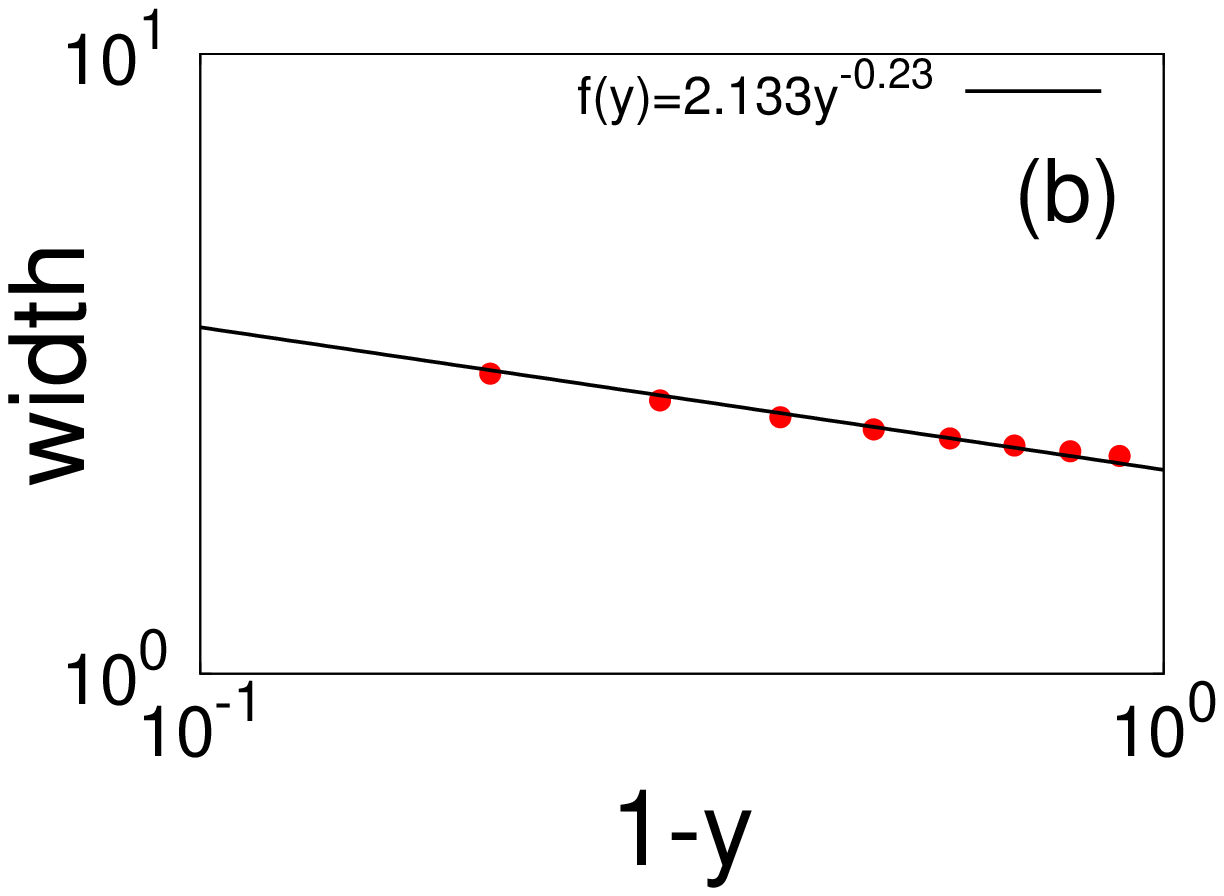}
 \caption{ Width of the Gaussian distributions for (a) $0.5 > z \geq 0; y=1$ and (b) $0 \leq  y < 1; z =0.5$ showing power 
law divergences.}
\label{width}
\end{figure}
\section {Distribution functions $S(x,t)$}

The behaviour of the probability distribution $S(x,t)$ for different values of $z$ and $y$ is discussed in this section. 

\subsection{$y = 1$}

The $y=1$ line shows different behaviour of $S(x,t)$ in the three regions $z< 0.5, ~ z = 0.5 $ and $z  > 0.5$, discussed separately
in the following subsections.

\subsubsection  {$0\leq z < 0.5$, $y=1$}

The  distributions $S(x,t)$ in this region are Gaussian 
in nature, just as one expects in a usual random walk
scenario. The scaling form obeyed by $S(x,t)$ for all values of $z$ between 0 and 0.5 is
\begin{equation}
S(x,t) \sim t^{-\frac{1}{2}}f_0(x/t^{\frac{1}{2}}),
\end{equation}
i.e. $\alpha=0.5$ (Fig. \ref{z0.5}(a)) and $f_0(X)=\exp(-\beta X^{2})$, where $\beta$ depends on $z$. 
While $x \sim t^{1/2}$ for the entire region, the  width
of the Gaussian form increases with $z$. It is possible to fit the widths  in the form $(0.5-z)^{-\mu_1}$ 
with $\mu_1 = 0.62$ (Fig. \ref{width}).

\subsubsection{$z=0.5$,$y =1$} 

This point which corresponds to the Voter model leads to the  most interesting results. 

Here $S(x,t)$
changes its behaviour markedly as time progresses. During early times, 
it shows a single peaked behaviour which, however, is non-Gaussian. 
For a system size of $128 \times 128$, the data collapse for different times (Fig. \ref{z0.5}b) could be obtained in the form given in Eq. (\ref{scaling})  with $\alpha \simeq 0.91.$
The scaling function is fitted to the form
\begin{equation}
f_{0.5}(X)=a[(b^2-X^2)]^c.
\label{f0.5}
\end{equation}
Keeping $b=1.75$, the best fit is obtained with
$a=0.201$ (with an error $\sim 10^{-4}$), and $c=1.38 \pm 0.002$. 
As $S(x,t)$ is vanishingly small beyond $X=1.75$, $b$ was kept fixed at 1.75.

\begin{figure}
\includegraphics[width=8cm]{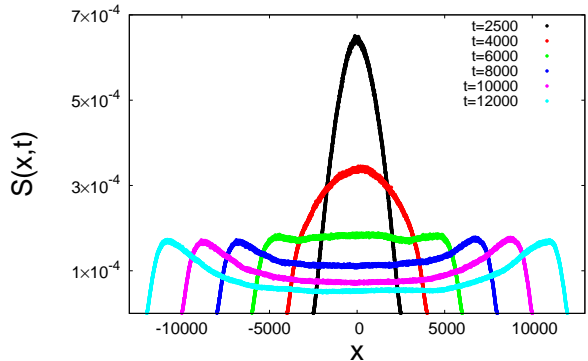}
 \caption{Crossover of the probability distribution $S(x,t)$ curves from a single peaked to a double peaked form for $z=0.5,y=1$ is shown for $L=64$. As time increases, the transition from single peaked form to double peaked form is shown.}
\label{crossover}
\end{figure}

\begin{figure}
\includegraphics[width=8cm]{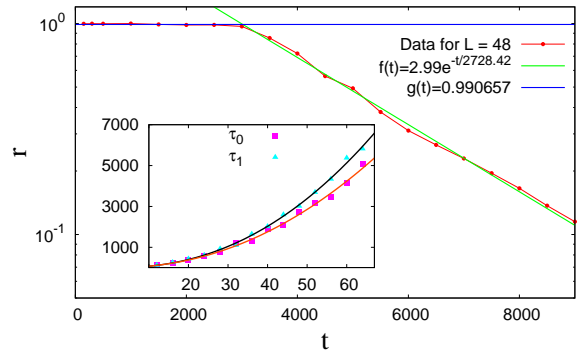}
 \caption{Variation of the ratio $r=\frac{S(x=0,t)}{S_{max}(t)}$ with time for $L=48$ system for $z=0.5,y=1$. Inset shows
the variation of $\tau_0$ and $\tau_1$ with system size $L$.}
\label{ratiovst}
\end{figure}

At larger timescales one can  observe a crossover from
the centrally peaked distribution to a double peak shaped distribution, where the
position of the two peaks are symmetric about the origin. The peaks occur at larger
values of $x$ (Fig. \ref{crossover}) as $t$ increases.

At large values of $t$, where the system is still not in equilibrium, the double peak
shaped curves cannot be fitted according to Eq. (\ref{scaling}) or any other simple form.
The transition from the single peaked form at earlier times to double peak shaped form at later times is gradual;
the peak at $x=0$ goes down while secondary peaks start growing with time. 

To characterize this
crossover behavior,  we estimate the ratio $ r = S(x=0,t)/S_{max}(t)$.
$S_{max}(t)$ is defined as the maximum value of the curve $S(x,t)$ ($x$ includes $0$).
A study of the variation of $r$ with time (Fig. \ref{ratiovst})
shows that it remains close to unity till a time $\tau_0$, then vanishes exponentially
for $t > \tau_0$. 
While $\tau_0$ is the timescale up to which the transient (single peaked) behaviour 
of $S(x,t)$ is observed,  another timescale $\tau_1$ may be defined as $r$ decreases
as $\exp(-t/{\tau_1})$ beyond $\tau_0$.


\begin{figure}
\includegraphics[width=8cm]{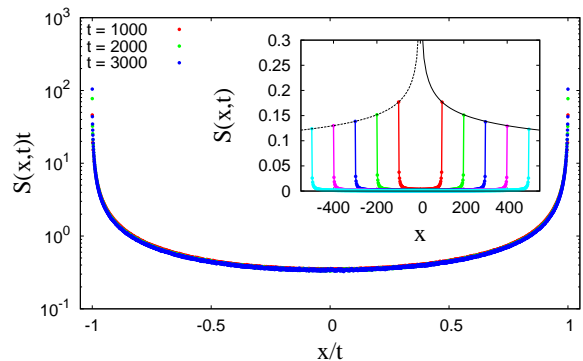}
 \caption{Data collapse for $S(x,t)$ for $z=0.8,y=1$ shown for $L=128$. Inset shows fitting of $S(x=t,t)$ curves for $z=1,y=1$
in the form of $\sim t^{-\theta_{2d}}$.}
\label{z0.8}
\end{figure}


The  study  of $\tau_0$ and $\tau_1$ with system size $L$ (inset of Fig. \ref{ratiovst}) shows that
both vary with system size $L$ as $L^2\log L$ to fairly good accuracy.
Precisely, $\tau_0 \sim L^{2.03\pm0.04}\log L$ and $\tau_1 \sim L^{1.91\pm0.066}\log L$.

\subsubsection{$z > 0.5, y=1$}

In this region one finds that $S(x,t)$ shows a `U' shaped behavior at all times and a 
data collapse can be obtained using Eq. (\ref{scaling}) with $\alpha =  1$.  Fig. \ref{z0.8} shows the case for $z=0.8$ as an example..
 For the Ising model ($z=1$), the `U' shaped distribution had been noted earlier \cite{drouffe2001}. 
Also, by fitting the tips of the distribution to the form $t^{ -\theta_{2d}}$ one can recover the persistence exponent $\theta_{2d} \approx 0.22$ 
(shown in the inset of Fig. \ref{z0.8}). For the intermediate region, $0.5 < z< 1.0$, the power law behaviour of persistence
can be obtained asymptotically only  \cite{drouffe1999,parna} and therefore  a fitting of the tips with a power law form does not work too well.

\subsection{$z = y$}

\begin{figure}
\includegraphics[width=8cm]{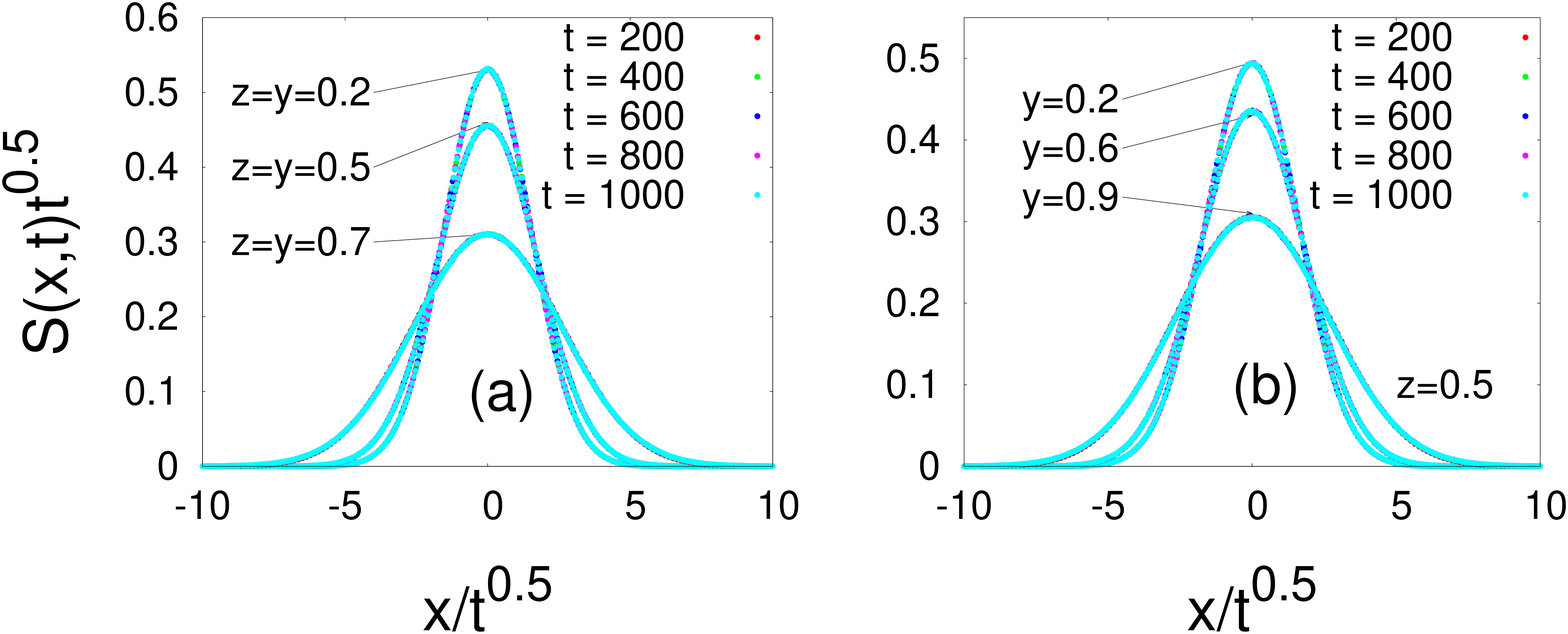}
 \caption{Data collapse of $S(x,t)$ for (a) $z=y=0.2,0.5,0.7$ and (b) $y=0.2,0.6,0.9;z=0.5$ for $L=128$. In both the cases, the collapse for a particular value of $z,y$ was performed taking five different times.}
\label{collapse_mv}
\end{figure}

The $z = y$ line corresponds to the majority vote model \cite{mv}.
Here $S(x,t)$ for walkers show a Gaussian form 
up to $z=y\approx 0.85$ (Fig. \ref{collapse_mv}a) and the width of the Gaussian function increases with $z$.

\begin{figure}
\includegraphics[width=8cm]{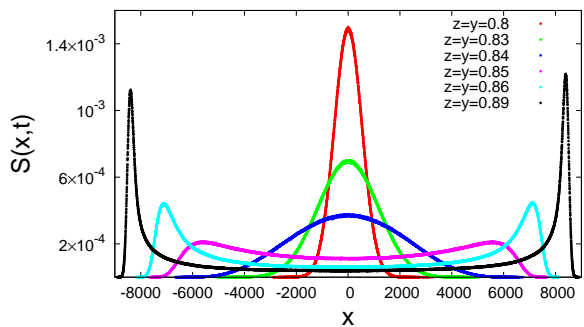}
 \caption{$S(x,t)$ curves obtained at $t=10^4$ for $L=128$ for different values of $z=y$. As the value of $z$ increases,
transition occurs from single peaked form to double peaked form.}
\label{crossover_mv}
\end{figure}

\begin{figure}
\includegraphics[width=8cm]{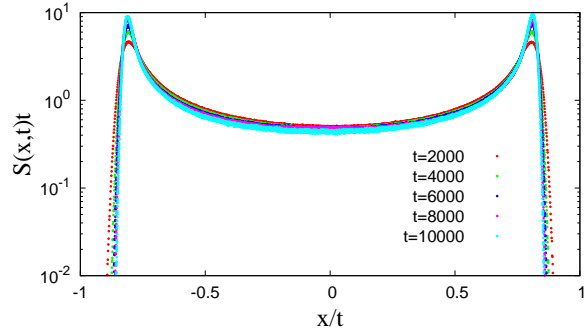}
 \caption{Approximate data collapse for $S(x,t)$ with $z=y=0.88$ for a system size of $128 \times 128$.}
\label{z0.88}
\end{figure}

For higher values of $z$, the Gaussian nature is replaced by
double peak shaped curves (Fig. \ref{crossover_mv}). An approximate  data collapse for $z=y>0.85$ using
Eq. (\ref{scaling}) with $\alpha =  1$ is obtained  in this region (Fig. \ref{z0.88}).

\subsection{$z = 0.5$, $y \geq 1$}

The probability distribution curves for the position of the walkers keeping $z=0.5$ and varying $y$ (Fig. \ref{collapse_mv}b)
shows a Gaussian scaling
form for $y\neq 1$. The width of the Gaussian function shows an increase with $y$.
Again, one finds a divergence of the width close to $y=1$ in the form $(1-y)^{-\mu_2}$ with $\mu_2 = 0.23$. 
The $y=1$ point is already discussed in section IIA.

\section{Analysis of the results}

\begin{figure}
\includegraphics[width=8cm]{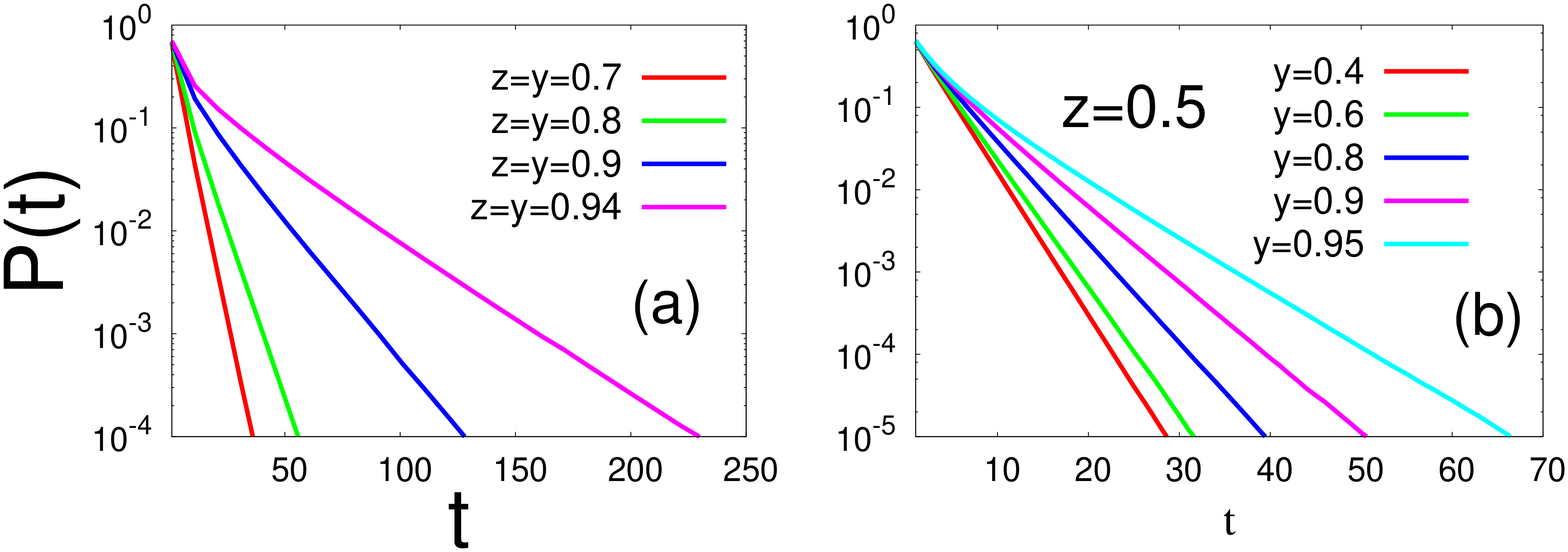}
 \caption{Persistence probability $P(t)$ as a function of time for a $256 \times 256$ system for (a) for four different values of $z=y$ and (b) for $z = 0.5$ with
five different values of $y$. The figures demonstrate exponential decay of $P(t)$ with time. The slopes of the curves decrease with increasing value of $y$.}
\label{per256}
\end{figure}

As mentioned in section I, we first check the consistency of $S(x,t)$ with the persistence probability.
Persistence probability $P(t)$ can be related to $S(x,t)$ as $S(x=t,t)=\frac{1}{2}P(t)$. From this, it follows that
systems for which $S(x,t)$ shows a Gaussian form,  i.e. $S(x,t)\sim \frac{1}{\sqrt{t}} \exp(\frac{-\beta x^2}{t})$, the persistence 
probability will have a form $P(t)\propto S(x=t,t)\sim \frac{1}{\sqrt{t}} \exp(-\beta t)$ which at large times is dominantly exponential.
The Gaussian behavior was seen for three regions; $z<0.5$ along the line $y=1$, $z=y\leq 0.85$ along the line
$z=y$ and for $z=0.5$ with $y<1$.
Indeed the persistence behavior in these particular cases show an exponential decay; some examples are shown in Fig. \ref{per256}.

The regions where $S(x,t)$ is non-Gaussian, 
one can still attempt to obtain the persistence behaviour by fitting $S(x=\pm t,t)$.   
For example, for $z=y=1$ where $S(t,t)$ has finite values, the known algebraic form $P(t) \sim t^{-\theta_{2d}}$ 
can be fit 
as shown in the inset of Fig. \ref{z0.8}.
For $z>0.5$ and $y=1$, 
$S(x,t)$ shows a finite value for $x= \pm t$ consistent 
with the  result that the persistence probability saturates to a constant value \cite{parna}. On the other hand,
for $z=y$, for $0.85<z<1$, where a non-Gaussian form is valid, we find that $S(x=t,t)$ is vanishingly small
 indicating  $P(t)$ here goes to zero much faster than a power
law decay. This is
checked by evaluating $P(t)$ directly (Fig. \ref{per256}). For the Voter model point $z=0.5, y=1$
also, $S(t,t)$ is comparatively much smaller than the Ising case, consistent with the fact
that persistence probability vanishes for the Voter model
as $P(t)\sim\exp[-a(\ln t)^2]$ \cite{bennaim,howard,maillard}.

A data collapse of $S(x,t)$ curves could be obtained according to Eq. (\ref{scaling})
for all cases except for the Voter model (beyond the transient time). The  values
of the exponent $\alpha$ are  either 0.5 corresponding to a diffusive walk
or (very  close) to 1 indicating a (nearly)  ballistic walk.

A diffusive walk arises when the spin flip probability is exactly 0.5. It can be easily shown that for a perfectly random configuration, the spin flip
probability is indeed 0.5 independent of $z$ and $y$. Hence a diffusive walk signifies that the system is in a disordered state. 
The dependence of the  parameter $\beta$ on $z$  signifies how fast is the diffusion; obviously $\beta$ is larger when the  system is `more' disordered (spin flip probabilities are closer to 1/2).

For $y=1$, when $z=0$, the spin flip probability is equal to 0.5 except for the fully aligned neighbourhood (all four neighbours in up/down state). Hence it is not surprising that 
the disordered state prevails for small $z$ leading to  a Gaussian form for $S(x,t)$.  
A  non-Gaussian form is obtained for $z \geq 0.5$. 
 
Along the line  $z = 0.5$, for $y <  1$, the spins can flip with  finite probabilities   
for all possible neighbourhoods, although the probability is not equal to $0.5$ in general. 
Thus  a disordered state is expected here and once again  $S(x,t)$ is found to be Gaussian.

We next discuss the results for the   $z=y$ line. 
For the limiting case $z=y=0$, the flipping probabilities for spins are 0.5 irrespective of the configurations and the system
will be disordered obviously. On the other hand, at $z = y =1$,  we have already observed a U shaped form for $S(x,t)$. 
   A change in the behaviour of $S(x,t)$ may occur 
either at some intermediate point or right at $z=1$.  
 Indeed, we find that it occurs at an intermediate point as for   $z>z_c \simeq 0.85$, $S(x,t)$ shows  a double peaked form beyond $z_c$.

We note a change in behaviour of $S(x,t)$ along all the three lines studied. 
We claim that this  change  from a Gaussian to a non-Gaussian form 
signifies the existence of an order-disorder transition as was found in some thermally driven phase transitions. 
For $z \geq 0.5, y=1$, it is already known that an ordered region exists 
\cite{oliveira,parna}. 
In fact, except for the point $z=y=1$ where frozen states occur 
with a finite probability, the ordered states are the  consensus states
(all up/ all down states).  For  
$y=1$,   
it is easy to see that the consensus state will be an absorbing state and hence the system will remain unchanged if it is reached, which happens at $z \geq 0.5$. 

For $z=y$, previously obtained results have shown that close to $z_c$, an order-disorder phase transition indeed exists \cite{oliveira,drouffe1999}.  
The present authors have also studied the behavior of magnetisation which shows that ordered states exists beyond $z_c$. 
Hence  the  change in behaviour of $S(x,t)$ 
indeed signifies   an  order disorder transition   consistent with earlier studies.

We next try to infer from the behaviour of $S(x,t)$ whether the 
ordered state is an active state. 
Taking  $\alpha = 1$  it has been possible to get a data collapse for $S(x,t)$ in the  region  $z > 0.5, y=1$ very accurately which  indicates it is an absorbing state agreeing with earlier studies \cite{parna}. 
On the other hand, for $z=y$, the data collapse with $\alpha=1$ beyond $z_c$ is only approximately valid.

\begin{figure}
\includegraphics[width=8cm]{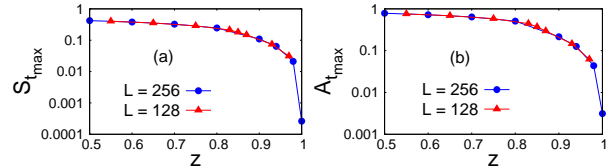}
 \caption{Values of (a) $S_t$ (fraction of spin flips) and (b) $A_t$ (fraction of active bonds) at a maximum time $t=10^6$ as a function of $z$ for two different system sizes for systems with $z=y$.}
\label{satval}
\end{figure}

It is easy to check that for $z = y \neq 1$, even when the system is fully ordered, 
the spins can flip with  a finite probability  equal to $(1-z)/2$.
A single  spin  flip will give rise to four active bonds 
(an active bond is defined as 
a pair of neighbouring spins with opposite orientation) in a uniform 
background of like spins. 
Indeed the results 
for the density of active bonds $A_t$ and spin flips $S_t$ as functions of time 
are consistent with this picture as even very close to $z=1$,  finite values of $A_t$ and $S_t$ exist, which have order of magnitude higher than that compared to $z=1$ (Fig. \ref{satval}). 
So the state is obviously an active state.  
However one cannot extract the information 
that there is an order-disorder transition at $z_c$ from the behaviour of 
$A_t$ and $S_t$ (there is no finite size effect also) while $S(x,t)$ clearly indicates it. 

\begin{figure}
\includegraphics[width=8cm]{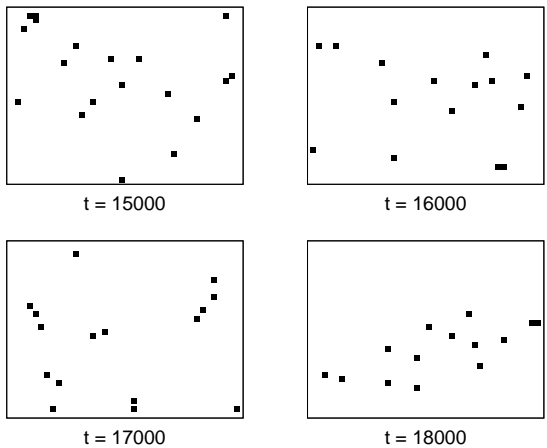}
 \caption{Snapshots at different times during the evolution of a $40\times 40$ system with $z=y=0.98$. The black squares
represent the down spins in the sea of up spins (white region).}
\label{snap}
\end{figure}

Although the system does not go to an absorbing state
for $z=y > z_c$, 
 since with $\alpha = 1$ 
an approximate data collapse can be obtained,  
it appears that most of the spins 
remain in their states for large times. 
The snapshots (Fig. \ref{snap}) support this picture as we find that 
there are random and rare spin flips continuing in the system causing the 
deviation from  a pure ballistic walk. 

Hence $S(x,t)$ successfully shows the existence of order disorder phase transitions 
along $y=1$ and $z=y$. 
Also, that the ordered state is an absorbing state for $z > 0.5, y=1$ is 
indicated by the fact that using $\alpha =1$ gives a much better 
scaling collapse for the curves compared to that for  $z = y \geq 0.85$ 
where an active state exists. 

One can also attempt to explain the crossover behavior of $S(x,t)$ in the Voter model.
At earlier times a single peaked non-Gaussian behavior has been observed.
Clearly, the system is not completely disordered so a Gaussian behavior is absent.
The peak at earlier times which occurs at the origin apparently signifies that a considerable
number of spins randomly flip till time $\tau_0$ although the system is
not completely disordered. Beyond $\tau_0$, ordering is the dominant phenomenon. It may be conjectured
that this crossover occurs when the surface noise becomes dominant over the bulk noise, a well studied phenomenon for the voter model \cite{hinrichsen}. Thus
the recurrent behavior of the walks disappears and $S(x,t)$ becomes double peaked at large times.

The width of the Gaussian distribution functions obtained along  both the lines $0 \leq z < 0.5$; $y=1$ and 
 $z = 0.5$,  $y <  1$  shows a power law divergence close to the Voter model point.
However, the divergences occur with different exponents; this may be connected to the   different roles played by   the two
noise factors also.

\section{Summary and Conclusions}

In this paper, a virtual walk in the so called spin space was  considered for 
a family of two dimensional classical spin models termed as generalised voter models.   The nature of the 
walks depends strongly on the parameters chosen but in general it could be fit 
to a form given by Eq. (\ref{scaling}). 
Either   $\alpha=0.5$ representing a diffusive walk or  $\alpha \simeq 1$ signifying a  (nearly) ballistic walk is obtained for the different regions in the parameter
space of the model. 

The voter model is an exception and showed a unique behavior
for which a universal scaling behavior could not be obtained using any simple form. 
Moreover, it  shows an interesting  crossover behaviour and the presence of two time scales. Interestingly, these two time scales show the same type of scaling as the consensus time of the Voter model. The fact that $\tau_0$, the transient time scales as $L^2\log L$ indicates that this crossover is not a finite size effect.
The Voter model can be mapped to a system of coalescing random walkers, and the existence of the crossover may be a reflection  of the fact that for random walks,
the upper critical dimension is two. 

We note that $S(x=t,t)$ is proportional to the
persistence probability and this was verified for several of the models. It was also found that  order-disorder transitions can be detected
from the change in behavior of $S(x,t)$. Further, whether the ordered state is an active state or an absorbing state
can be conjectured.

In conclusion, the coarsening dynamics for some classical spin models have been studied
regarding spins as walkers in a manner similar to some earlier studies.    
 It is possible to extract important information from the distribution of the distance travelled by the 
walker which is  related to the difference in time spent with different spin states. 
The connection with  persistence behaviour is obvious.
Order disorder transition  in certain cases can be detected more efficiently 
 compared to other dynamical variables.  Even in the disordered region, power law divergences in the 
width are noted which appear with nontrivial exponents.

%




Acknowledgement: The authors are grateful to Claude Godr\`eche for comments on the manuscript and for drawing attention to 
several earlier studies. P. Mullick thanks DST-INSPIRE (Sanction No. 2015/IF0673) for their financial support.
P. Sen acknowledges SERB (Government of India) for their financial grant.


\begin{thebibliography}{99}

\bibitem{derrida1}
B. Derrida, A. J. Bray and C. Godr\`eche, 1994 J. Phys. A \textbf{27}, L357 (1994).

\bibitem{derrida2}
B. Derrida, J. Phys. A \textbf{28}, 1481 (1995); B. Derrida, V. Hakim, and V. Pasquier, Phys. Rev. Lett. \textbf{75}, 751 (1995).

\bibitem{Bray1}
A. J. Bray, Adv. Phys. \textbf{51}, 481 (2002).

\bibitem{Bray2}
A. J. Bray, S. N. Majumdar and G. Schehr, Adv. Phys. \textbf{62}, 225 (2013).

\bibitem{parna}
P. Roy and P. Sen, Phys. Rev. E \textbf{95}, 020101(R) (2017).

\bibitem{ligget}
T. M. Liggett, \textit{Interacting Particle Systems} (Springer, New York, 1985).


\bibitem{dornic_g98}
I. Dornic and C. Godr\`eche, J. Phys. A \textbf{31} 5413 (1998).

\bibitem{drouffe98}
J. M. Drouffe and C. Godr\`eche, J. Phys. A \textbf{31}, 9801 (1998).

\bibitem{newman}
T. J. Newman and Z.Toroczkai, Phys. Rev. E \textbf{58}, R2685 (1998).

\bibitem{balda}
A. Baldassarri, J. P. Bouchaud, I. Dornic and C. Godr\`eche Phys. Rev. E \textbf{59}, R20 (1999).

\bibitem{luck}
C. Godr\`eche and J. M. Luck, J. Stat. Phys. \textbf{104}, 489 (2001).

\bibitem{peng}
C. K. Peng et. al., Nature \textbf{356}, 168 (1992).

\bibitem{bachelier}
L. Bachelier, \textit{Theorie de la speculation}, Annales Scientifiques de I'Ecole Normale Superiure, \textbf{3} (17), pp. 21-86 (1900).

\bibitem{econo1}
A. Chatterjee and P. Sen, Phys. Rev. E \textbf{82}, 056117 (2010).

\bibitem{econo2}
S. Goswami, A. Chatterjee and P. Sen, Phys. Rev. E \textbf{84}, 051118 (2011).

\bibitem{drouffe2001}
J. M. Drouffe and C. Godr\`eche, Eur. Phys. J. B \textbf{20}, 281 (2001).

\bibitem{oliveira}
M. J. de Oliveira, J. F. F. Mendes and M. A. Santos, J. Phys. A Math. Gen. \textbf{26}, 2317 (1993).

\bibitem{drouffe1999}
J. M. Drouffe and C. Godr\`eche, J. Phys. A \textbf{32}, 249 (1999).

\bibitem{mv}
M. J. de Oliveira, J. Stat. Phys. \textbf{66}, 273 (1992).

\bibitem{bennaim}
E. Ben-Naim, L. Frachebourg and P. L. Krapivsky, Phys. Rev. E \textbf{53}, 3078 (1996).

\bibitem{howard}
M. Howard and C. Godr\`eche, 1998 J. Phys. A \textbf{31}, L209 (1998).

\bibitem{maillard}
G. Maillard and T. Mountford, Annales de l'Institut Henri Poincar\'e Probabilit\'es et Statistiques \textbf{45}, 577 (2009).






\bibitem{hinrichsen}
I. Dornic, H. Chat\'e, J. Chave and H. Hinrichsen, Phys. Rev. Lett. \textbf{87}, 045701 (2001).




%


\end{thebibliography}
\end{document}